\documentclass[aps,showpacs,prb,twocolumn]{revtex4}
\usepackage{mathrsfs}
\usepackage{amssymb}
\usepackage{amsmath}
\usepackage{bm}
\usepackage{graphics}
\usepackage{natbib}

\begin{document}
\title{Detection of spin bias in four-terminal quantum-dot ring}

\author{Weijiang Gong$^{1,3}$}
\author{Hui Li$^{1}$}
\author{Sha Zhang$^{1}$}
\author{Yu Han$^{2}$}
\author{Guozhu Wei$^{1,3}$}

\affiliation{ 1. College of Sciences, Northeastern University,
Shenyang 110004, China \\
2. Department of Physics, Liaoning University, Shenyang 110036,
China\\
3. International Center for Material Physics, Acadmia Sinica,
Shenyang 110015, China}
\date{\today}

\begin{abstract}
In this work, we show that in a four-quantum-dot ring, via
introducing a local Rashba spin-orbit interaction the spin bias in
the transverse terminals can be detected by observing the charge
currents in the longitudinal probes. It is found that due to the
Rashba interaction, the quantum interference in this system becomes
spin-dependent and the opposite-spin currents induced by the spin
bias can present different magnitudes, so charge currents emerge.
Besides, the charge currents rely on both the magnitude and spin
polarization direction of the spin bias. It is believed that this
method provides an electrical but practical scheme to detect the
spin bias (or the spin current).

\end{abstract}
\pacs{73.63.Kv, 73.21.La, 73.23.Hk, 85.35.Be} \maketitle

\bigskip

\section{Introduction}
The field of spintronics has become a significant concern of both
experimental and theoretical communities for the purpose of
realizing the quantum information processing.\cite{Wolf,Dasarma}
Under such a topic, researchers devoted themselves to the
fabrication of various nano-devices to efficiently generate and
manipulate the spin current by magnetic means including the
application of strong magnetic field or ferromagnetic electrodes to
achieve the spin injection,\cite{Sato,Nature1} or by the optical
approach consisting of employing the polarized
light.\cite{Sipe1,Sipe2,Cui,Other} Recently, the Rashba spin-orbit
(SO) interaction is recommended to realize the spin manipulation in
the low-dimensional structures, because it offers a electric manner
to control the electron spin.\cite{Rashba1,Rashba2,Rashba3} Albeit
these existed works, on the other hand, the quantitative detection
of spin current is now at its infancy, because that the occurrence
of the pure spin current is usually not accompanied by any electric
signal for direct measurement.\cite{detect}
\par
As mentioned in the previous literature, there have been a variety
of proposals to focus on the measurement of spin current. Kato
\emph{et al.} first experimentally observed the spin Hall current
via the magneto-optical Kerr effect in GaAs semiconductor
systems.\cite{kato} Other suggestions to detect spin current include
measuring the induced mechanical deformation of the macroscopic
sample or measuring the induced spin torque.\cite{spintorque}
Recently, a report showed that the spin current can be detected in a
double-QD system when the presented Coulomb interaction destroys the
symmetry of spin-up and spin-down current.\cite{Sun} However, the
schemes described above are comparatively complicated, thus any
simple one to achieve the detect the spin current is still
desirable.
\par
It is distinct that for a spin current flowing through a structure,
a spin-dependent chemical potential (spin bias) is usually induced
that is the driving force of spin current, thus we can measure the
spin bias instead of spin current. With such an idea, in this work
by adopting the local Rashba SO interaction we suggest to
electrically measure the spin bias in virtue of the quantum
interference in a QD ring. Its key point is that the local Rashba
interaction gives rise to the spin-dependent quantum interference
and the electron transmission through the QD ring is then
spin-polarized. Therefore, the spin polarization in this QD ring
breaks the symmetry of the motion of spin-up and spin-down
electrons, and nonzero spin-bias--induced charge current
correspondingly come up.

\section{model and formulation\label{theory}}
The considered four-QD ring is illustrated in Fig.\ref{Structure}.
We assume that in the transverse leads ( lead-1 and -3 ) there exist
the spin bias $V_{s}$, i.e., the spin-dependent chemical potentials
for the spin-up and spin-down electrons are
$\mu_{1\sigma}=\mu_{3\bar{\sigma}}=\varepsilon_F+\sigma eV_{s}$ with
the Fermi level of the system $\varepsilon_F$ and $\sigma=\pm1$ (or
$\uparrow, \downarrow$) being the spin index. We additionally insert
two normal metallic probes ( lead-2 and -4 ) longitudinally to
observe the electric signal change of them affected by the spin
bias, so as to ascertain the existence of spin bias. The
single-electron Hamiltonian in this structure can be written as
$H_s=\frac{\textbf{P}^2}{2m^*}+V(\textbf{r})+\frac{\hat{y}}{2\hbar}\cdot[\alpha(\hat{\sigma}\times
\textbf{p})+(\hat{\sigma}\times \textbf{p})\alpha]$, where the
potential $V(\textbf{r})$ confines the electron to form the
structure geometry, namely, the leads, QDs and the connections. The
last term in $H_s$ denotes the local Rashba SO coupling on QD-2. For
the analysis of the electron properties, we have to second-quantize
the Hamiltonian,\cite{gongapl} which is composed of three parts:
${\cal H}={\cal H}_{c}+{\cal H}_{d}+{\cal H}_{t}$.
\begin{eqnarray}
{\cal H}_{c}&&=\underset{\sigma jk}{\sum }\varepsilon
_{jk\sigma}c_{jk\sigma}^\dag c_{jk\sigma },\notag\\
{\cal H}_d&&=\sum_{j=1, \sigma}^{4}\varepsilon
_{j}d^\dag_{j\sigma}d_{j\sigma}
+\sum_{l=1,\sigma}^{2}[t_{l\sigma}d^\dag_{l\sigma}d_{l+1\sigma}+r_l(d_{l\downarrow}^\dag
d_{l+1\uparrow}\notag\\&&-d_{l+1\downarrow}^\dag
d_{l\uparrow})]+t_{3}d^\dag_{3\sigma}d_{4\sigma}+t_{4}e^{i\phi}d^\dag_{4\sigma}d_{1\sigma}+\mathrm
{H.c.},\notag\\ {\cal H}_{t}&&=\underset{\sigma jk }{\sum
}V_{j\sigma} d^\dag_{j\sigma}c_{jk\sigma}+\mathrm {H.c.},
\end{eqnarray}
where $c_{jk\sigma}^\dag$ and $d^{\dag}_{j\sigma}$ $( c_{jk\sigma}$
and $d_{j\sigma})$ are the creation (annihilation) operators
corresponding to the basis in lead-$j$ and QD-$j$. $\varepsilon
_{jk\sigma}$ and $\varepsilon_{j}$ are the single-particle levels.
$V_{j\sigma}$ denotes QD-lead coupling strength. The interdot
hopping amplitude is written as
$t_{l\sigma}=t_l\sqrt{1+\tilde{\alpha}^2}e^{-i\sigma\varphi}$
($l=1,2$) where $t_{l}$ is the ordinary transfer integral
independent of the Rashba interaction and $\tilde{\alpha}$ is the
dimensionless Rashba coefficient with
$\varphi=\tan^{-1}\tilde{\alpha}$\cite{Serra}. The phase factor
$\phi$ attached to $t_{4}$ accounts for the magnetic flux through
the ring. In addition, the many-body effect can be readily
incorporated into the above Hamiltonian by adding the Hubbard term
${\cal
V}_{e\text{-}e}=\sum_{j\sigma}{\frac{U_j}{2}}n_{j\sigma}n_{j\bar{\sigma}}$.
\par
Starting from the second-quantized Hamiltonian, we can now formulate
the electronic transport properties. With the nonequilibrium Keldysh
Green function technique, the current flow in lead-$j$ at the zero
temperature can be written as\cite{Meir}
\begin{equation}
J_{j}=\frac{e}{h}\sum_{j'\sigma\sigma'}\int_{\mu_{j'\sigma'}}^{\mu_{j\sigma}}
d\omega T_{j\sigma,j'\sigma'}(\omega),\label{current}
\end{equation}
where $T_{j\sigma,j'\sigma'}(\omega)=4\Gamma_ {j\sigma}
 G^r_{j\sigma,j'\sigma'}(\omega)\Gamma_{j'\sigma'}G^a_{j'\sigma',j\sigma}(\omega)$
is the transmission function, describing electron tunneling ability
between lead-$j$ to lead-$j'$. $\Gamma_{j\sigma}=\pi
|V_{j\sigma}|^2\rho_j(\omega)$, the coupling strength between QD-$j$
and lead-$j$, can be usually regarded as a constant.  $G^r$ and
$G^a$, the retarded and advanced Green functions, obey the
relationship $[G^r]=[G^a]^\dag$. From the equation-of-motion method,
the retarded Green function can be obtained in a matrix form,
\begin{widetext}
\begin{eqnarray}
&&[G^r]^{-1}=\notag\\
&&\left[\begin{array}{cccccccc} g_{1\uparrow}^{-1} & -t_{1\uparrow}&0&-t_{4}e^{-i\phi}&0&r^*_1&0&0\\
  -t^*_{1\uparrow}& g_{2\uparrow}^{-1}& -t_{2\uparrow}&0&-r^*_1&0&r^*_2&0\\
  0&-t^*_{2\uparrow}&g_{3\uparrow}^{-1}&-t_{3}&0&-r^*_2&0&0 \\
  -t_{4}e^{i\phi}&0&-t^*_{3}&g_{4\uparrow}^{-1}&0&0&0&0 \\
  0&-r_1&0&0&g_{1\downarrow}^{-1}&-t_{1\downarrow}&0&-t_{4}e^{-i\phi}\\
  r_1&0&-r_2&0&-t^*_{1\downarrow}& g_{2\downarrow}^{-1}& -t_{2\downarrow}&0\\
  0&r_2&0&0&0&-t^*_{2\downarrow}&g_{3\downarrow}^{-1}&-t_{3}\\
  0&0&0&0&-t_{4}e^{i\phi}&0&-t^*_{3}&g_{4\downarrow}^{-1}
\end{array}\right]\notag.
\end{eqnarray}
\end{widetext}
In the above expression, $g_{j\sigma}$ is the Green function of
QD-$j$ unperturbed by the other QDs and in the absence of Rashba
effect.
$g_{j\sigma}=[(z-\varepsilon_{j})\lambda_{j\sigma}+i\Gamma_j]^{-1}$
with $z=\omega+i0^+$ and
$\lambda_{j\sigma}=\frac{z-\varepsilon_{j}-U_{j}}{z-\varepsilon_{j}-U_{j}+U_j\langle
n_{j\bar{\sigma}}\rangle}$ resulting from the second-order
approximation of the Coulomb interaction\cite{Zhenghz}. $\langle
n_{j\sigma}\rangle$ can be numerically resolved by the formula
$\langle n_{j\sigma}\rangle=-\frac{i}{2\pi}\int d\omega
G^{<}_{j\sigma,j\sigma}$ where
$G^<_{\sigma\sigma}=\sum_{\sigma'}[G^r]_{\sigma\sigma'}[\Sigma^<]_{\sigma'}[G^a]_{\sigma'\sigma}$
and $[\Sigma^<]_\sigma=2\sum_j\Gamma_{j\sigma}f_{j\sigma}(\omega)$.
$f_{j\sigma}(\omega)=\theta(\mu_{j\sigma}-\omega)$ is the Fermi
distribution function in lead-$j$ with the step function of
$\theta(x)$.

\section{Numerical results and discussions \label{result2}}
\par
We now proceed on to calculate the currents in the longitudinal
terminals, lead-2 and lead-4 in this case. Before calculation, the
QD-lead couplings are assumed to take the uniform values with
$\Gamma_{j\sigma}=\Gamma$, and we consider $\Gamma$ as the energy
unit ( Its order is \emph{meV} approximately for some experiments
based on GaAs/GaAlAs QDs, as mentioned in the previous
works\cite{PRL,PRB} ). The structure parameters are for simplicity
taken as $|t_{l\sigma}|=t_3=t_4=\Gamma$, and $\varepsilon_F$ is
viewed as the energy zero point of this system. Besides, to carry
out the numerical calculation, we choose the Rashba coefficient
$\tilde{\alpha}=0.4$ which is available in the current
experiment.\cite{Sarra}
\par
We first show the linear-transport results. It is known that in the
linear regime, the current flow is proportional to the linear
conductance, i.e., $J_{m}=\mathcal {G}_{m}\cdot V_s$ ($m=1,3$),
where the linear conductance
\begin{eqnarray}
\mathcal {G}_{m}=\frac{e^{2}}{h}[&&\bar{\sigma}
(T_{m\sigma,1\sigma}+T_{m\sigma,3\bar{\sigma}})\notag\\
&+&\sigma(T_{m\sigma,1\bar{\sigma}}+
T_{m\sigma,3\sigma})]|_{\omega=\varepsilon_F } \label{conduct}
\end{eqnarray}
obeys the Landauer-B\"{u}ttiker formula.\cite{Datta} Consequently,
in this case, by only investigating the properties of the linear
conductance the spin-bias driven charge current can clarified. From
such a formula, one can readily find that in the absence of any
spin-dependent fields the electron transmission is irrelevant to the
electron spin. Hence the opposite-spin currents driven by the spin
bias flow through this ring with the same magnitude and opposite
directions, leading to the result of zero ${\cal G}_{m}$ and the
failure of measuring the spin bias [ see the dashed line in
Fig.\ref{Cond}(a)].
\par
On account of the recent researches\cite{ee}, they show that in the
low-dimensional systems, such as the QD structures, in the electron
transport process introducing a local Rashba interaction could bring
out the spin polarization, which helps to manipulate the electron
spin via the electric means. Thereupon, we introduce a local Rashba
interaction to QD-2 of this structure to try detecting the spin bias
in the transverse leads. As shown in Fig.\ref{Cond}(a), in the
presence of Rashba SO coupling and the absence of magnetic field,
there indeed emerge the nonzero currents in the longitudinal probes
driven by the spin bias when the QD levels are separate from the
energy zero point (i.e., $\varepsilon_0\neq 0$). An additional
interesting phenomenon is that in the whole regime the amplitude of
$J_{2}$ is the same as that of $J_{4}$ accompanied by their opposite
directions. Such a result means that by building a closed circuit
between lead-2 and -4 the spin bias of this system can be detected
by observing the current flowing between the longitudinal probes.
Besides, it is found that the direction of charge current is related
to the separation of QD levels from the Fermi level, namely, in the
case of $\varepsilon_0>0$ the value of $J_{2}$ is less than zero and
$J_{4}>0$, whereas $J_{2}>0$ and $J_{4}<0$ under the condition of
$\varepsilon_0<0$.
\par
Since the configuration of quantum ring, we would like to
investigate the role of a local magnetic flux. Thereby, we can see
that for the case of $\phi={\pi\over 2}$ there appears little
spin-bias-induced currents in the probes despite the adjustment of
QD levels, as shown in Fig.\ref{Cond}(b). Alternatively, when the QD
levels take a typical value with $\varepsilon_0=\Gamma$, the
currents present finite values except in the vicinity of
$\phi=(n+{1\over 2})\pi$, and they oscillate out of phase.
Significantly, as shown in Fig.\ref{Cond}(c) the change of magnetic
flux from $\phi=2n\pi$ to $\phi=(2n+1)\pi$ can vary the magnitude
and direction of the spin-bias-driven charge currents in the
longitudinal terminals with the critical point at $\phi
=(2n+\frac{1}{2})\pi$, and vise versa. Up to now, we can address
that the presented Rashba interaction and the nonzero QD levels with
respect to the zero point of energy are the two key conditions to
accomplish the measure of the spin bias electrically in this
structure.
\par
For the case of the finite spin bias, the charge currents in the
additional terminals can be evaluated by Eq.(\ref{current}).
Accordingly, in Fig.\ref{Bias}(a) we plot the current spectra
\emph{vs} the QD levels in the situations of $eV_s=\Gamma$ and
$2\Gamma$, respectively. One can find that in such a case the
current magnitudes increase with the enhancement of the spin bias.
And the current spectra exhibit complicated properties, different
from the linear-transport case. For the case of
$|\varepsilon_0|>{V_s\over 2}$, the quantitative relation between
these two charge currents becomes ambiguous, especially in the case
of $|\varepsilon_0|> 2\Gamma$ the signs of these two charge currents
can be the same as each other. Only when the QD levels shift around
the Fermi level of the system ( i.e., in the regime of
$|\varepsilon_0|<{V_s\over 2}$) the result of $J_2=-J_4$ remains
substantially. Similarly, such a phenomenon is also described by
Fig.\ref{Bias}(b). As a typical case, when taking the QD levels at
$|\varepsilon_0|=\Gamma$, we see that in the situation of
$eV_s<\Gamma$ the charge currents in the longitudinal probes have
the same magnitude and the opposite directions. However, with the
enhancement of the spin bias the value of $J_4$ goes over the zero
point and then shows the identical sign with $J_2$.

\par
The calculated transmission functions are plotted in Fig.\ref{Trans}
with $\varepsilon_j=\Gamma$. They are just the integrands for the
calculation of the charge and spin currents (see
Eq.(\ref{current})). By comparing the results shown in
Figs.\ref{Trans}(a), we can readily see that in the absence of
magnetic flux, the traces of $T_{2\uparrow,1\uparrow}$,
$T_{4\downarrow,1\downarrow}$, $T_{2\downarrow,3\downarrow}$, and
$T_{4\uparrow,3\uparrow}$ coincide with one another very well, so do
the curves of $T_{2\downarrow,1\downarrow}$,
$T_{4\uparrow,1\uparrow}$, $T_{2\uparrow,3\uparrow}$, and
$T_{4\downarrow,3\downarrow}$. Substituting such integrands into the
current formulae, one can certainly arrive at the result of the
distinct pure spin currents in the transverse terminals. On the
other hand, these transmission functions depend nontrivially on the
magnetic phase factor, as exhibited in Figs.\ref{Trans}(b) with
$\phi={\pi\over 2}$. In comparison with the zero magnetic field
case, herein the spectra of $T_{j\uparrow,j'\uparrow}$ are reversed
about the axis $\omega=\Gamma$ without the change of their
amplitudes, but $T_{j\downarrow,j'\downarrow}$ only present the
enhancement of their amplitudes. Similarly, with the help of
Eq.(\ref{current}), one can understand the disappearance of spin
currents in such a case. In addition, by virtue of the transmission
function curves we can understand the behaviors of the charge
currents with the enhancement of spin bias, i.e., when the strength
of spin bias goes beyond the quantum coherence regime the current
feature displayed in the linear regime disappears.
\par
The underlying physics being responsible for the spin dependence of
the transmission functions is quantum interference, which manifests
if we analyze the electron transmission process in the language of
Feynman path. Note that the spin flip terms arising from the Rashba
interaction do not play a leading role in causing the appearance of
spin and charge currents\cite{Gong-APL}. Therefore, to keep the
argument simple, we drop the spin flip terms for the analysis of
quantum interference. Based on this method, we write
$T_{2\sigma,1\sigma}=|\tau_{2\sigma,1\sigma}|^2$ where the
transmission probability amplitude is defined as
$\tau_{2\sigma,1\sigma}=\widetilde{V}^*_{2\sigma}
G^r_{2\sigma,1\sigma}\widetilde{V}_{1\sigma}$ with
$\widetilde{V}_{j\sigma}=V_{j\sigma}\sqrt{2\pi\rho_j(\omega)}$. With
the solution of $G^r_{2\sigma,1\sigma}$, we find that the
transmission probability amplitude $\tau_{2\sigma,1\sigma}$ can be
divided into three terms, i.e.,
$\tau_{2\sigma,1\sigma}=\tau^{(1)}_{2\sigma,1\sigma}+\tau^{(2)}_{2\sigma,1\sigma}
+\tau^{(3)}_{2\sigma,1\sigma}$, where
$\tau^{(1)}_{2\sigma,1\sigma}=\frac{1}{D}\widetilde{V}^*_{2\sigma}
g_{2\sigma}t^*_{1\sigma}g_{1\sigma}\widetilde{V}_{1\sigma}$,
$\tau^{(2)}_{2\sigma,1\sigma}=\frac{1}{D}\widetilde{V}^*_{2\sigma}
g_{2\sigma}t_{2\sigma}g_{3\sigma}t_{3}g_{4\sigma}t_4e^{i\phi}
g_{1\sigma}\widetilde{V}_{1\sigma}$, and
$\tau^{(3)}_{2\sigma,1\sigma}=-\frac{1}{D}\widetilde{V}^*_{2\sigma}
g_{2\sigma}t_{2\sigma}g_{3\sigma}t^*_{2\sigma}g_{2\sigma}t^*_{1\sigma}
g_{1\sigma}\widetilde{V}_{1\sigma}$ with
$D=\det\{[G^r]^{-1}\}\prod_jg_{j\sigma}$. By observing the
structures of $\tau^{(1)}_{2\sigma,1\sigma} $,
$\tau^{(2)}_{2\sigma,1\sigma} $, and $\tau^{(3)}_{2\sigma,1\sigma}$,
we can readily find that they just represent the three paths from
lead-2 to lead-1 via the QD ring. The phase difference between
$\tau^{(1)}_{2\sigma,1\sigma}$ and $\tau^{(2)}_{2\sigma,1\sigma}$ is
$\Delta\phi^{(1)}_{2\sigma}=[\phi-2\sigma\varphi+\theta_3+\theta_4]$
with $\theta_j$ arising from $g_{j\sigma}$, whereas the phase
difference between $\tau^{(2)}_{2\sigma,1\sigma}$ and
$\tau^{(3)}_{2\sigma,1\sigma}$ is
$\Delta\phi^{(2)}_{2\sigma}=[\phi-2\sigma\varphi]$. It is clear that
only these two phase differences are related to the spin
polarization. $T_{4\sigma,1\sigma}$ can be analyzed in a similar
way. We then write
$T_{4\sigma,1\sigma}=|\tau^{(1)}_{4\sigma,1\sigma}+\tau^{(2)}_{4\sigma,1\sigma}+\tau^{(3)}_{4\sigma,1\sigma}|^2$,
with $\tau^{(1)}_{4\sigma,1\sigma}
=\frac{1}{D}\widetilde{V}^*_{4\sigma}g_{4\sigma}t_{4}e^{i\phi}g_{1\sigma}\widetilde{V}_{1\sigma}
$,
$\tau^{(2)}_{4\sigma,1\sigma}=\frac{1}{D}\widetilde{V}^*_{4\sigma}g_{4\sigma}
t^*_3g_{3\sigma}t^*_{2\sigma}g_{2\sigma}t^*_{1\sigma}g_{1\sigma}\widetilde{V}_{1\sigma}
$, and
$\tau^{(3)}_{4\sigma,1\sigma}=-\frac{1}{D}\widetilde{V}^*_{4\sigma}g_{4\sigma}
t^*_{3}g_{3\sigma}t_{3}g_{4\sigma}t_4e^{i\phi}g_{1\sigma}\widetilde{V}_{1\sigma}
$. The phase difference between $\tau^{(1)}_{4\sigma,1\sigma}$ and
$\tau^{(2)}_{4\sigma,1\sigma}$ is
$\Delta\phi^{(1)}_{4\sigma}=[\phi-2\sigma\varphi-\theta_2-\theta_3]$,
and $\Delta\phi^{(2)}_{4\sigma}=[\phi-2\sigma\varphi]$ originates
from the phase difference between $\tau^{(2)}_{4\sigma,1\sigma}$ and
$\tau^{(3)}_{4\sigma,1\sigma}$. Utilizing the parameter values in
Fig.\ref{Trans}, we evaluate that $\varphi\approx{\pi\over 7}$ and
$\theta_j=-\frac{3\pi}{4}$ at the point of $\omega=0$. It is
apparent that when $\phi=0$ only the phase differences
$\Delta\phi^{(1)}_{2\sigma}$ and $\Delta\phi^{(1)}_{4\sigma}$ are
spin-dependent. Accordingly, we obtain that
$\Delta\phi^{(1)}_{2\uparrow}=-\Delta\phi^{(1)}_{4\downarrow}={3\pi\over
14}$, and
$\Delta\phi^{(1)}_{2\downarrow}=-\Delta\phi^{(1)}_{4\uparrow}={11\pi\over
14}$, which clearly prove that the quantum interference between
$\tau^{(1)}_{2\uparrow,1\uparrow}$ and
$\tau^{(2)}_{2\uparrow,1\uparrow}$
($\tau^{(1)}_{4\downarrow,1\downarrow}$ and
$\tau^{(2)}_{4\downarrow,1\downarrow}$ alike) is constructive, but
the destructive quantum interference occurs between
$\tau^{(1)}_{2\downarrow,1\downarrow}$ and
$\tau^{(2)}_{2\downarrow,1\downarrow}$
($\tau^{(1)}_{4\uparrow,1\uparrow}$ and
$\tau^{(2)}_{4\uparrow,1\uparrow}$ alike). Then such a quantum
interference pattern can explain the traces of the transmission
functions shown in Fig.\ref{Trans}(a). In the case of
$\phi=\frac{\pi}{2}$ we find that only
$\Delta\phi^{(2)}_{2(4)\sigma}$ are crucial for the occurrence of
spin polarization. By a calculation, we obtain
$\Delta\phi^{(2)}_{2(4)\uparrow}={5\pi\over 14}$ and
$\Delta\phi^{(2)}_{2(4)\downarrow}={9\pi\over 14}$, which are able
to help us clarify the results in Fig.\ref{Trans}(c) and (d). Up to
now, the characteristics of the transmission functions, as shown in
Fig.\ref{Trans}, hence, the tunability of charge currents have been
clearly explained by analyzing the quantum interference between the
transmission paths.

\par
So far we have not discussed the effect of electron interaction on
the occurrence of charge currents in the longitudinal probes, though
it is included in our theoretical treatment. Now we incorporate the
electron interaction into the calculation, and we deal with the
many-body terms by employing the second-order approximation, since
we are not interested in the electron correlation here.
Fig.\ref{Coulomb} shows the calculated currents spectra with
$U_j=U=3\Gamma$, respectively. Clearly, within such an approximation
the spin-bias-induced charge currents remain, though the current
spectra oscillate to a great extent with the shift of QD levels, as
shown in Fig.\ref{Coulomb}(a). On the other hand, the numerical
results in Fig.\ref{Coulomb}(b) tell us that when the QD levels are
aligned with the zero point of energy of this structure, by the
presence of Coulomb-interaction the charge currents is possible to
appear with the further increase of the applied bias, different from
those in the noninteracting case. Meanwhile, in the case of the QD
levels separate from the energy zero point ($\varepsilon_j=\Gamma$),
the magnitudes of the charge currents seem to be enhanced by the
many-body effect. This can also be understood with the help of the
discussion on the quantum interference of this structure above.

\section{Summary}
\par

In conclusion, when introducing a local Rashba interaction on an
individual QD of a four-QD ring, we proposed to electrically detect
the spin bias of the transverse terminals by investigating the
charge currents in the two external longitudinal probes. We have
found that the quantum interference in this system becomes
spin-dependent by the presence of the Rashba interaction, so the
opposite-spin currents driven by the spin bias show different
magnitudes, leading to the emergence of the charge currents.
Besides, the charge currents rely on both the magnitude and spin
polarization direction of the spin bias. Therefore, this method
provides a practical and electrical approach to detect the spin
bias. On the other hand, the modulation of the QD levels and the
magnetic phase factor can efficiently adjust the phases of the
transmission paths, so properties of the spin bias can be shown
entirely. Finally, it should be emphasized that altering the
polarization directions of the spin bias, equivalent to interchange
the sequence numbers of lead-1 and lead-3, can also change the
polarization directions of the charge currents.

\section*{Acknowledgments}

This work was financially supported by the National Natural Science
Foundation of China (Grant No. 10904010), the Seed Foundation of
Northeastern University of China (Grant No. N090405015), and the
Scientific Research Project of Liaoning Education Office (Grant No.
2009A309).

\clearpage

\bigskip

\begin{figure}
\caption{ (a) Schematic of a
four-QD ring structure with a local Rashba interaction on QD-2. Four
QDs and the leads coupling to them are denoted as QD-$j$ and
lead-$j$ with $j=1-4$. Spin bias is assumed to be in lead-2 and
lead-4. \label{Structure}}
\end{figure}

\begin{figure}
\caption{ In (a) and (b), the linear conductances vs the QD levels
are shown with the magnetic phase factor $\phi=0$ and $\pi\over 2$,
respectively. The parameter values are $\Gamma_j=\Gamma$ and
$\tilde{\alpha}=0.4$. (c) The linear conductance vs $\phi$ with
$\varepsilon_j=\Gamma$. \label{Cond}}
\end{figure}

\begin{figure}
\caption{ The charge currents in the case of finite spin bias. (a)
The currents versus as functions of the QD levels with $eV_S=\Gamma$
and $2\Gamma$, respectively. (b) The change of the currents with the
adjustment of spin bias strength.\label{Bias}}
\end{figure}

\begin{figure}
\caption{ The spectra of transmission functions
$T_{m\sigma,m'\sigma}$($m$=2,4 and $m'$=1,3) with the QD levels
fixed at $\varepsilon_j=\Gamma$. (a) and (b) Zero magnetic field
case, and (c)-(d) magnetic phase factor $\phi={\pi\over
2}$.\label{Trans}}
\end{figure}

\begin{figure}
\caption{ In the presence of many-body terms with $U_j=3\Gamma$, the
currents versus $\varepsilon_0$ and the spin bias strength,
respectively. The other parameters are the same as those in
Fig.\ref{Bias}.\label{Coulomb}}
\end{figure}

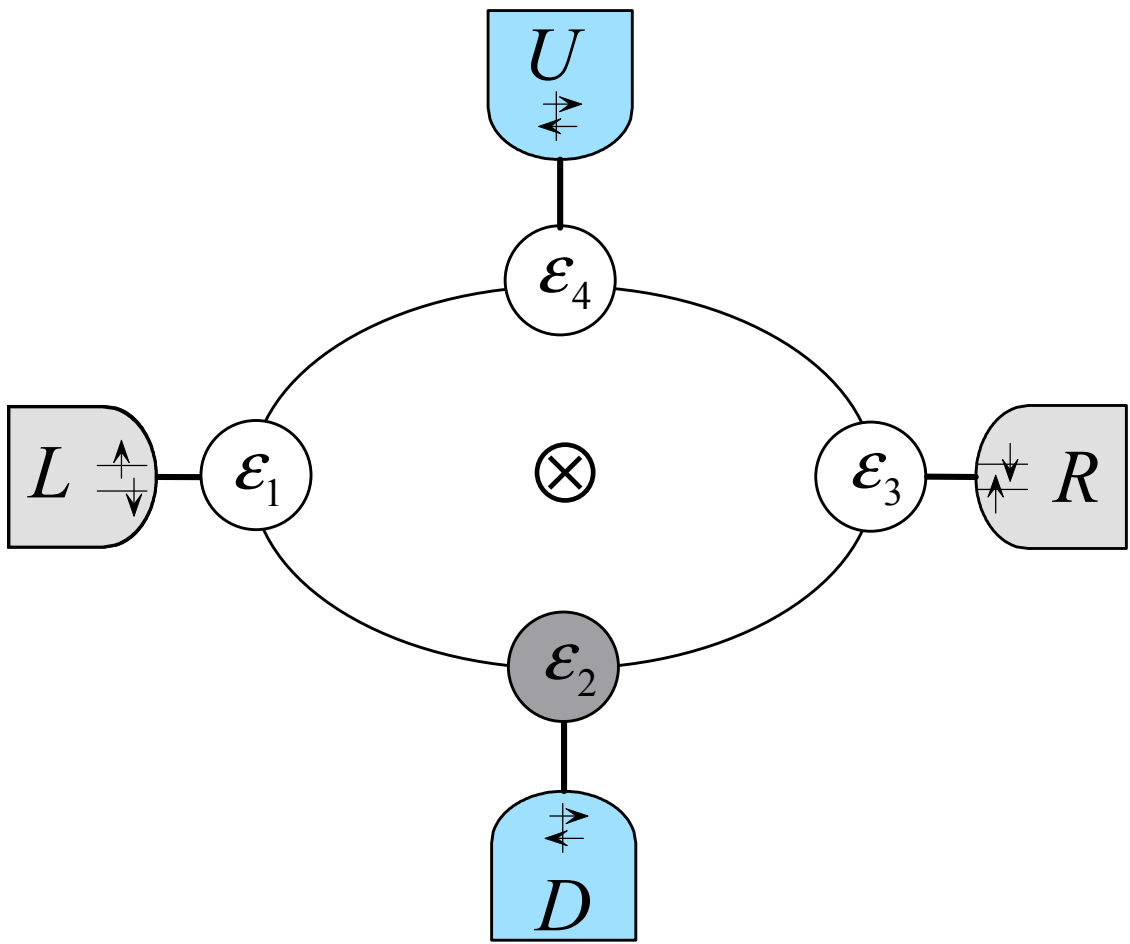
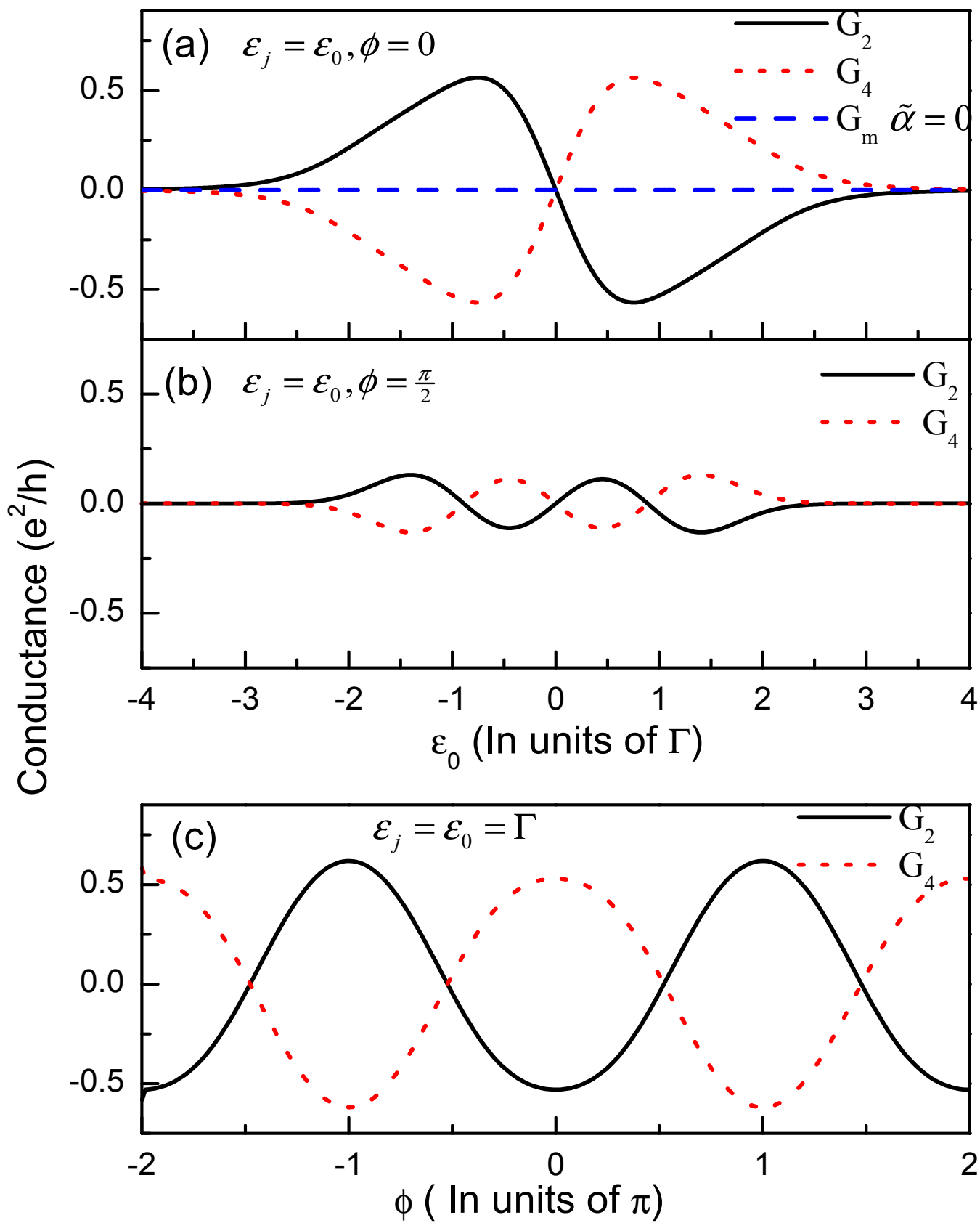
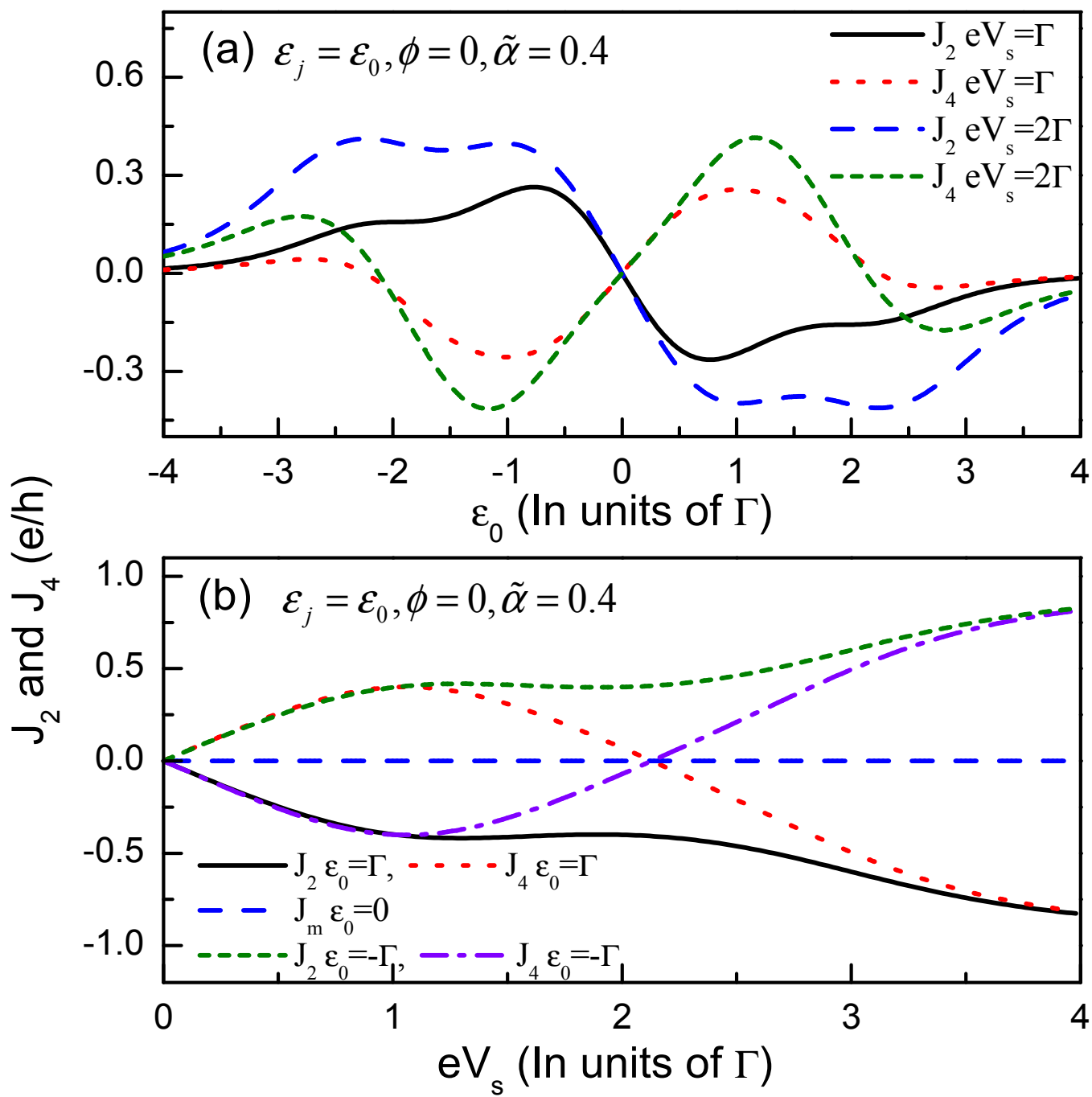
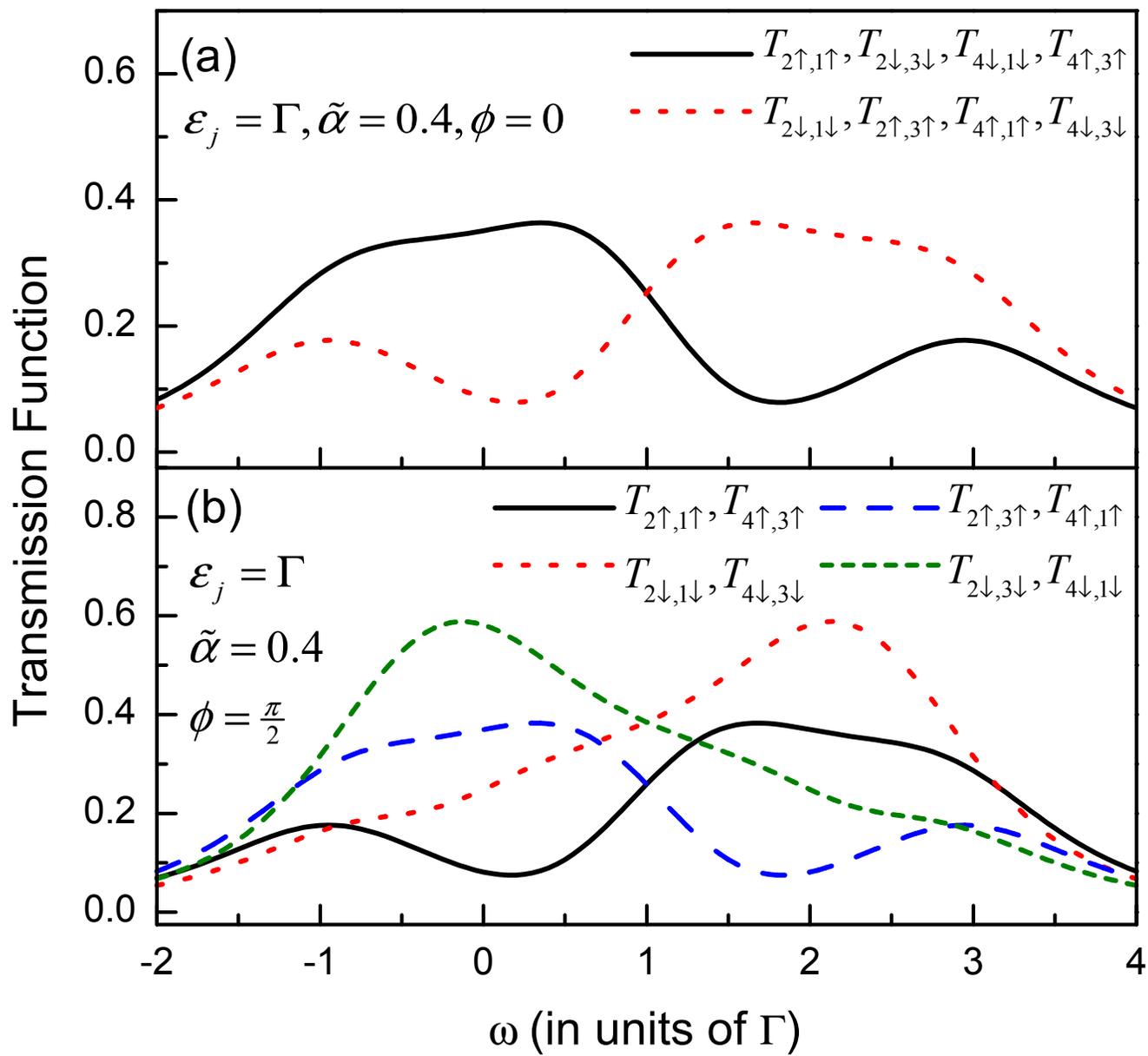
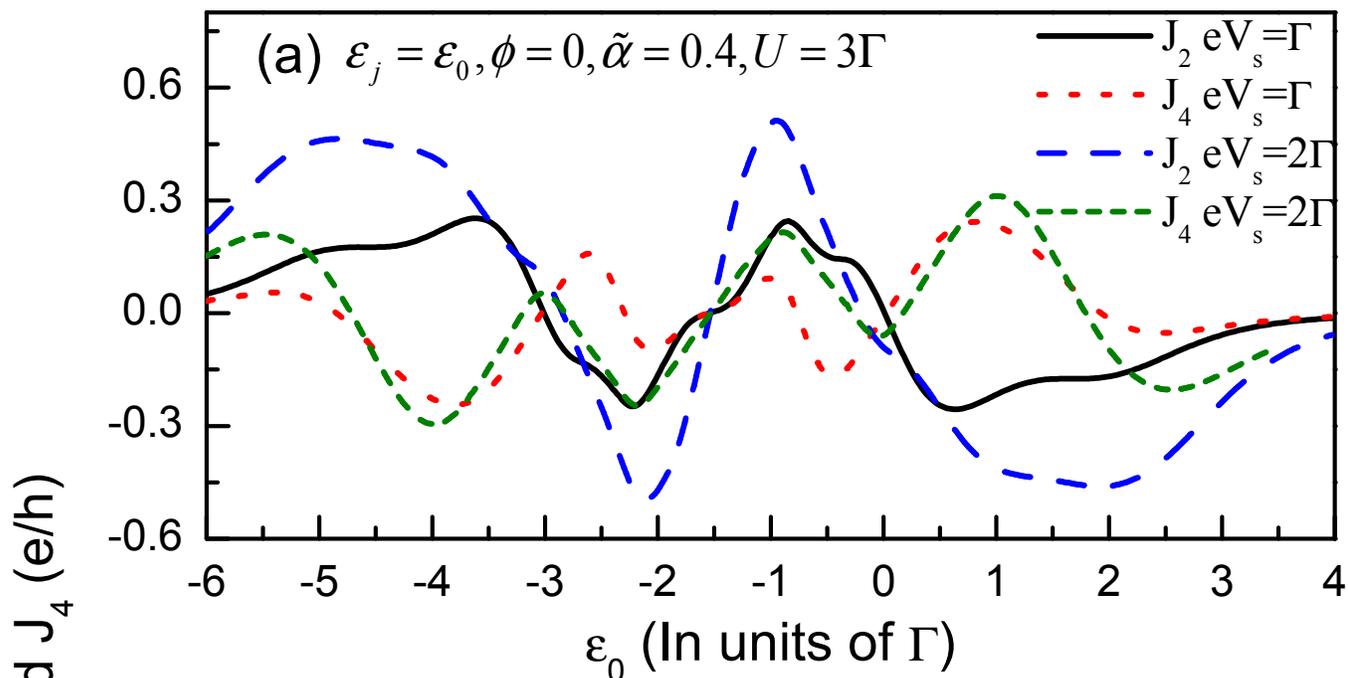
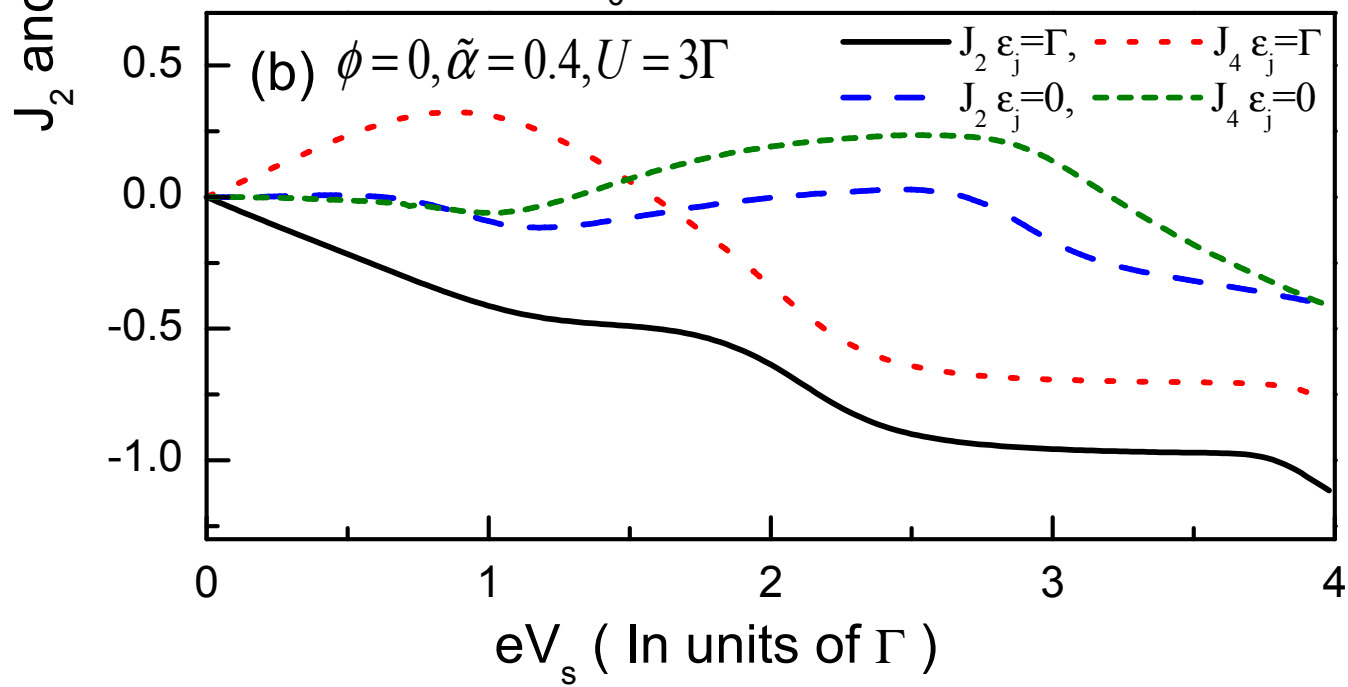
\end{document}